\documentstyle{article}

\newcommand{\ds}{\displaystyle}
\newcommand{\bea}{\begin{eqnarray}}
\newcommand{\eea}{\end{eqnarray}}
\newcommand{\beq}{\begin{equation}}
\newcommand{\eeq}{\end{equation}}
\newcommand{\tfr}{R\!\!\!\!\!\!\!\nearrow}
\newcommand{\tft}{T\!\!\!\!\!\!\!\nearrow}
\newcommand{\tsBrule}{%
            \pagebreak[3]\noindent\rule[.5ex]{\textwidth}{0.3mm}\hfill
            \vspace{0.5ex}}
\newcommand{\tsErule}{%
            \noindent\rule[.5ex]{\textwidth}{0.3mm}\hfill\pagebreak[3]}
\newcommand{\Paw}{Paw{\l}owski}
\newcommand{\Rac}{R\c{a}czka}
\newcommand{\ts}{$\;$}
\begin{document}
\title{A Quadratic Curvature Lagrangian\\ 
       of \Paw\ and \Rac:\\
       A Finger Exercise with MathTensor%
\thanks{To appear in Hehl, F.W., Puntigam, R.A., Ruder, H. (eds.): 
       {\em Relativity and Scientific Computing -- Computer Algebra, 
       Numerics, Visualization. Lectures given at the $152^{\rm th}$
       WE-Heraeus Seminar, 18--22 Sept 1995, Bad Honnef (Germany).}
       Springer, Berlin, 1996}}
\author{Efstratios Tsantilis, Roland A. Puntigam, 
        and Friedrich W.\ Hehl\\[2mm]
{\small Institute for Theoretical Physics, University of Cologne, Germany}\\
{\small  e-mail: {\tt Hehl@ThP.Uni-Koeln.De}}}
\maketitle
\begin{abstract}
Recently \Paw\ and \Rac\ (P\&R) proposed a unified 
model for the fundamental interactions which does not contain a 
physical Higgs field. \index{Higgs field} 
The {\it gravitational} field equation of 
their model is rederived under heavy use of the computer algebra 
system Mathematica and its package MathTensor.
\end{abstract}
\section{Introduction} 

The computer algebra system Mathematica\index{Mathematica}, 
together with its \index{MathTensor}
package MathTensor, is very useful for executing differential 
geometric calculations on a four-dimensional Riemannian 
spacetime, as was shown by Soleng \cite{soleng}, for example. Here we 
want to demonstrate the effectiveness of these tools by picking 
an example from the current literature which is of some 
fundamental importance. \par

The gravitational sector of a unified model for the fundamental%
\index{standard model}\index{P\&R model}
interactions, proposed by P\&R \cite{PW}, is determined by the
Lagrangian density 
\beq 
   {\cal L}_{\rm geom}= \sqrt{-g}\left( -{1\over
    6}\,(1+\beta)\,R\Phi^\dagger\Phi -\lambda(\Phi^\dagger\Phi)^2
  -\rho\, C_{ijkl}\,C^{ijkl} \right)\,,\label{lgeo} 
\eeq 
with $\Phi$ as Higgs field and $R$ (curvature scalar) and 
$C_{ijkl}$ (conformal Weyl curvature tensor) as gravitational fields
depending on the metric tensor $g_{ij}$ of the Riemannian spacetime,
$g:=\det(g_{ij})$.\index{Weyl tensor}\index{conformal!invariance}
The coupling constants $\beta$, $\lambda$, and $\rho$ are
dimensionless because of the conformal invariance of the model.

The total Lagrangian density of P\&R,
\beq 
    {\cal L}= {\cal L}_{\rm geom} + {\cal L}_{\rm field}\;, \label{ltot}
\eeq 
depends additionally on the field part, which subsumes the
contributions from all nongeometrical pieces. The explicit form of
${\cal L}_{\rm field}$ is of no relevance to us\footnote{From the
Higgs-type Lagrangian, two pieces feature in the geometrical
Lagrangian, namely the first two terms in (\ref{lgeo}), whereas the
rest of it is attributed here to the field Lagrangian in
(\ref{ltot}).} since it is absorbed in the definition of the
`material' energy--momentum tensor%
\index{energy-momentum tensor!`material'}
\beq
   T^{ij}:={2\over\sqrt{-g}}\,{\delta{\cal L}_{\rm field} \over\delta
   g_{ij}}\,.\label{emt} 
\eeq 
Hamilton's principle yields the gravitational field 
equation\index{Hamilton's principle} 
\beq
   {\delta{\cal L}_{\rm geom}\over\delta g_{ij}} +
   {\delta{\cal L}_{\rm field}\over\delta g_{ij}}=0\label{feq1} \;,
\eeq
or
\beq
-{1\over\sqrt{-g}}\,{\delta{\cal L}_{\rm geom}\over\delta 
g_{ij}}= {1\over 2}\, T^{ij}\,.\label{feq2}
\eeq
The total Lagrangian (\ref{ltot}) is
conformally invariant.  Therefore it is possible, within the P\&R
model, to fix a conformal gauge for the scalar Higgs-type field $\Phi$
according to 
\beq 
   \Phi^\dagger\,\Phi={v^2\over 2}={\rm constant}\,.\label{fix} 
\eeq
Our goal will be the computation of the left-hand side of 
(\ref{feq2}) by means of the computer algebra tools 
mentioned above. Thereby we want to check the corresponding 
results of P\&R \cite{PW}.

\section{Riemann Tensor and its Irreducible Pieces}

Before we can commence with our calculations, we have to get hold 
of the definition of the Weyl curvature tensor $C_{ijkl}$. It is 
known that the curvature tensor of a four-dimensional Riemannian 
space  has three irreducible pieces: the Weyl tensor $C_{ijkl}$, 
the tracefree Ricci tensor $\tfr_{ij}$, and the curvature scalar 
$R$.\index{Ricci tensor!trace-free}\index{curvature scalar}\par

We take the conventions of Misner et al.\ \cite{MTW} and define 
the curvature tensor (${\Gamma^{i}}_{jk} =$ Christoffel symbols) 
\beq 
   {R^{i}}_{jkl} := \partial_{k} {\Gamma^{i}}_{jl}-\partial_{l} 
                    {\Gamma^{i}}_{jk}+ {\Gamma^{i}}_{mk} \, {\Gamma^{m}}_{jl}-
                    {\Gamma^{i}}_{ml} \, {\Gamma^{m}}_{jk} \,,\label{riem}
\eeq
the Ricci tensor and the curvature scalar
\beq
   R_{ij}:=R^k{}_{ikj}\,\qquad {\rm and}
   \qquad R:=g^{ij}\,R_{ij}\,,\label{ricci}
\eeq
respectively, and, eventually, the tracefree Ricci tensor
\beq
   \tfr_{ij}:=R_{ij}-{1\over 4}\,g_{ij}\,R\,.\label{tracefreericci}
\eeq
The metric has signature $(-+++)$.\par

The Riemann tensor (20 independent components) decomposes into three 
irreducible pieces, $20=10\oplus 9\oplus 1$, or%
\index{Riemann tensor!irreducible  decomposition of}
\beq
R_{ijkl}=\,^{(1)}R_{ijkl}+\,^{(2)}R_{ijkl}+\,^{(3)}R_{ijkl}\,,
\label{irrdecomp}
\eeq
with the definitions
\beq
   ^{(2)}R_{ij}{}^{kl}:=2\tfr_{[i}{}^{\![k}\,\delta_{j]}^{l]}\qquad
   {\rm and}\qquad ^{(3)}R_{ij}{}^{kl}:={1\over 6}\,R\,\delta_{\;i}^{[k}\,
   \delta_j^{l]}\,.\label{irrpieces}
\eeq
The first irreducible piece is traceless and has to be identified 
with the Weyl piece: $C_{ijkl} := \,^{(1)}R_{ijkl}$. If resolved with 
respect to $C_{ijkl}$, Eq.\ts (\ref{irrdecomp}) can be read as the 
defining equation for the Weyl tensor.

Let us use MathTensor in order to perform the irreducible decomposition%
\index{MathTensor}
defined in (\ref{irrdecomp}) and (\ref{irrpieces}). MathTensor
recognizes {\tt TraceFreeRicciR}. Therefore 
the pieces $^{(a)} R_{ijkl}$, denoted by  {\tt R1}, {\tt R2}, and {\tt R3},
can be computed by an almost verbatim translation of  
(\ref{irrdecomp}) and (\ref{irrpieces}) into MathTensor:

\tsBrule
\begin{verbatim}Dimension=4; Rcsign=1; (*Default*)
DefUnique[TraceFreeRicciR[la_,lb_],0,PairQ[la,lb]] 
R2[li_,lj_,lk_,ll_]=2 Antisymmetrize[Antisymmetrize[
     TraceFreeRicciR[lk,li] Metricg[lj,ll],{li,lj}],{lk,ll}]
R3[li_,lj_,lk_,ll_]=Expand[Antisymmetrize[
         1/6 ScalarR Metricg[lk,li] Metricg[lj,ll],{li,lj}]]
R1[li_,lj_,lk_,ll_]=RiemannR[li,lj,lk,ll]
                    - R2[li,lj,lk,ll] - R3[li,lj,lk,ll]
\end{verbatim}
\tsErule
We shall verify this decomposition with MathTensor.  
We first use the predefined tensor {\tt WeylC} in order to check if the
definitions are correct.  

Furthermore we want to make sure that our
decomposition is really irreducible. For this purpose we compute the
traces of $^{(a)} R_{ijkl}$, and find -- as well as the usual 
symmetries of the curvature tensor -- that $C_{ijkl}$ is traceless, 
the Ricci tensor of $^{(2)}R_{ijkl}$ is $\tfr_{ij}$, and 
$^{(3)}R_{ijkl}$ has neither a traceless piece nor a traceless Ricci 
piece, rather only the curvature scalar with the correct factor 
$1$. The input reads: 

\tsBrule
\begin{verbatim}TraceFreeRicciR[li,lj]/.TraceFreeRicciToRicciRule

diff1=CanAll[(WeylC[li,lj,lk,ll]-R1[li,lj,lk,ll])
              /.WeylToRiemannRule]
diff2=CanAll[Tsimplify[CanAll[diff1
              /.TraceFreeRicciToRicciRule]]]

TraceR2=Expand[Metricg[ua,ub] R2[la,li,lb,lj]]
TraceR3=Expand[R3[la,li,lb,lj]  Metricg[ua,ub]]
TraceR1=Tsimplify[Expand[Metricg[ua,ub] R1[la,li,lb,lj]
              /.TraceFreeRicciToRicciRule]]\end{verbatim}
\tsErule

\section{The Topological Euler Density}

Let us come back to (\ref{lgeo}). Its last term is proportional to 
\beq
   {\cal C}^2:=\sqrt{-g}\,C_{ijkl}\,C^{ijkl}\,.\label{c^2} 
\eeq
Densities will be denoted by script letters. The computation of the
Bach tensor \cite{bach}\index{Bach tensor}
\beq 
   B^{ij}:={1\over\sqrt{-g}}\,{\delta\,{\cal C}^2\over
   \delta g_{ij}}\label{bach}
\eeq
(see also \cite{schim}) can be simplified, if one splits off 
a divergence term from (\ref{c^2}).\index{Euler density}
Such a term is the topological Euler density \cite{egg}
\beq
   {\cal E}:=-{1\over 128\pi^2}\,\sqrt{-g}\,\varepsilon^{abcd}R^{ij}{}_{ab}
   \,R^{kl}{}_{cd}\,\varepsilon_{ijkl}\,.\label{euler}
\eeq
We can show that (\ref{euler}) represents a divergence:
\beq
   {\cal E}=\partial_i\left(\frac{1}{128\pi^2}{\cal D}^i\right)
\eeq
with
\beq
   {\cal D}^i:= \sqrt{-g} \,\varepsilon^{ijkl}\,
    {\varepsilon_{ab}}^{cd} \,  {\Gamma^{a}}_{cj}
    \left(\frac{1}{2} {R^{b}}_{dkl}+ \frac{1}{3}\, {\Gamma^{b}}_{mk}\,
    {\Gamma^{m}}_{dl}\right) \,.  \label{div}
\eeq
Furthermore, also by means of MathTensor, the explicit form of 
(\ref{euler}) turns out to be 
\beq 
 {\cal  E}:=\frac{\sqrt{-g}}{32 \pi^2} \, (\,R_{ijkl}\,R^{ijkl}-4 \,R_{ij}\, 
R^{ij}+R^2)
\,.   \label{euler1}
\eeq
The corresponding input reads:

\tsBrule
\begin{verbatim}Eulerd1=
     -1/(128 Pi^2) Epsilon[ua,ub,uc,ud] Epsilon[li,lj,lk,ll]
                RiemannR[ui,uj,la,lb] RiemannR[uk,ul,lc,ld]
Eulerd2=CanAll[Expand[Eulerd1/.EpsilonProductTensorRule]]\end{verbatim}
\tsErule
We substitute into (\ref{c^2}) the definition of the Weyl tensor and
eliminate the emerging curvature square piece by means of (\ref{euler1}). 
Then we find the simplified formula
\beq 
   {\cal C}^2= \check{{\cal C}}^2 +\partial_i{\cal D}^i\, ,
\eeq
with
\beq
   \check{{\cal C}}^2 :=  2\,(\,R^{ij}R_{ij}\,-\, \frac{R^2}{3})=
    2\,(\,\tfr ^{ij}\tfr _{ij}\,-\, \frac{R^2}{12})
   \,, \label{ccheck^2}
\eeq 
which has also been cross-checked by means of MathTensor:

\tsBrule
\begin{verbatim}WeylSquare1=WeylC[li,lj,lk,ll] WeylC[ui,uj,uk,ul]
WeylSquare2=CanAll[Dum[WeylSquare1/.WeylToRiemannRule]]
checkCsquare=Expand[WeylSquare2 - 32 Pi^2 Eulerd2]
checkCsquare2=Expand[checkCsquare
                     /.RicciToTraceFreeRicciRule]\end{verbatim}
\tsErule

\section{Bach Tensor}

In order to compute the variation of the remaining piece $\check{C}^2$
of the Lagrangian, we follow closely the scheme that was demonstrated in the
lecture by Soleng \cite{soleng}. After initalization, we use 
{\tt Variation} in order to compute $\delta\, \check{C}^2$, 
followed by a series of partial integrations, rules, and simplifications.
The last step actually computes the variational derivative:%
\index{variational derivative}

\tsBrule
\begin{verbatim}varC1=Variation[Sqrt[-Detg]*checkCsquare,Metricg]
varC2=PIntegrate[varC1,Metricg]
varC3=PIntegrate[varC2,Metricg]
varC4=Canonicalize[Absorbg[ApplyRules[varC3,RiemannRules]]]
Bach1[ui_,uj_]=Tsimplify[VariationalDerivative[Expand[
                varC4/Sqrt[-Detg]],Metricg,li,lj]]
MetricgFlag=True\end{verbatim}
\tsErule
The outcome of this variation reads (a semicolon denotes the covariant 
derivative)
\begin{eqnarray} 
   \mbox{\tt Bach1}\cong B_{ij}&=&\frac{2}{3} \, R_{;i;j} -  2\,{R_{ij;k}}^{;k}
   +\frac{1}{3}\,g_{ij}\,{R_{;k}}^{;k} \nonumber \\
   &&-\frac{1}{3}\, R^2 g_{ij}  + \frac{4}{3} \, R \, R_{ij} + g_{ij} \,
   R_{kl}\, R^{kl} - 4 \, R^{kl} \, R_{ikjl} \; ,\label{varcheckC}
\end{eqnarray} 
\beq 
    B_{ij}=B_{ji}\qquad{\rm and}\qquad g^{ij}B_{ij}=B^k{}_k=0\,.\label{sym}
\eeq 
The result (\ref{varcheckC}) differs slightly from that of P\&R.%
\footnote{Their result reads, see  \cite[Eq.\ts (7.2)]{PW2} and 
          \cite[Eq.\ts (5.2)]{PW}:
          \begin{eqnarray} \label{pawvar}
             && -\frac{2}{3} \, R_{;i;j} +  2\,{R_{ij;k}}^{;k}
               - \frac{{\bf 2}}{3}\,g_{ij}\,{R_{;k}}^{;k} \nonumber \\
             &&  -\frac{1}{3}\, R^2 g_{ij} + \frac{4}{3} \, R \, R_{ij}
               + g_{ij} \, R_{kl}\, R^{kl} - 4 \, R^{kl} \, R_{ikjl}\nonumber
      \,.     \end{eqnarray} 
           The third term of the first line carries an incorrect factor 
two. Therefore their Bach tensor is no longer traceless, as is required
by conformal invariance. Up to a (conventional?) sign, 
their first line is identical 
to that of (\ref{varcheckC}), whereas the second lines coincide.}

In order to compute the Bach tensor (\ref{varcheckC}), we could  
have used (\ref{c^2}) rather than (\ref{ccheck^2}). 
In this case (\ref{varcheckC}) would pick up two additional terms
that compensate each other, as is explicitly shown in \cite{parker}.
We feel, however, that the present detour, via $\check{{\cal C}}^2$,
pays off in conceptual and computational simplicity.
\section{The Bach Tensor Streamlined}

The Bach tensor takes on a more transparent form if we express the
curvature pieces in (\ref{varcheckC}) exclusively in terms of the
irreducible pieces.\index{Bach tensor!irreducible decomposition of}

By default, MathTensor does not recognize that the trace
of $C_{ijkl}$ vanishes over abritrary indices. 
Therefore we explicitly define rules to take care of this fact.
After this preparatory 
step, we can directly reformulate the Bach tensor.
Thus we put in:

\tsBrule
\begin{verbatim}DefUnique[WeylC[la_,lb_,lc_,ld_],0,PairQ[la,lb] || 
          PairQ[la,lc] || PairQ[la,ld] || PairQ[lb,lc] || 
          PairQ[lb,ld] || PairQ[lc,ld]]

Bach2[li_,lj_]=Tsimplify[CanAll[Expand[Bach1[li,lj]/.
            RiemannToWeylRule/.RicciToTraceFreeRicciRule]]]\end{verbatim}
\tsErule
This computation yields
\begin{eqnarray}
\mbox{\tt Bach2}\cong B_{ij}&\!=\!&
                  \frac{2}{3}\, R_{;i;j}-\frac{1}{6}{R_{;k}}^{;k}\, 
                   g_{ij}-2\, {\tfr_{ij;k}}^{;k} \nonumber \\
                && +\frac{2}{3} R \, \tfr_{ij}
                   +4\, \tfr_{ki} {\tfr_j}^k 
                   -g_{ij}\, \tfr_{kl}  \tfr^{kl} -4\, \tfr^{kl} 
                   C_{ikjl} \,.
\label{irredvarC}
\end{eqnarray}
If we introduce 
 \beq
  {\gamma_{ij}}^{kl}:=\delta_i^{(k}\delta_j^{l)}
   -{1\over 4} \,g_{ij}\,g^{kl}\,,
 \eeq 
 where
 \beq
   \gamma_{[ij]}{}^{kl}=\gamma_{ij}{}^{[kl]}\equiv 0
   \qquad{\rm and} \qquad g^{ij}\gamma_{ij}{}^{kl}\equiv 0\,,
 \eeq
then the tracelessness $B^k{}_k=0$ and the symmetry $B_{ij} = B_{ji}$
of the Bach tensor become manifest:
\begin{eqnarray}
 B_{ij} & = &   \frac{2}{3}\,{\gamma_{ij}}^{kl}\,R_{;k;l}
              - 2\, \tfr_{ij;k}{}^{;k} \nonumber \\
        &   & + \frac{2}{3} \, R \, \tfr_{ij}
              + 4\,{\gamma_{ij}}^{kl}\, \tfr_{mk}\, {\tfr^{m}}_l 
                  -4\, \tfr^{kl} \, 
                   C_{ikjl} \; .
   \label{bach2}
\end{eqnarray}
The trick for the corresponding computation is again the use
of a series of rules that expresses the Bach tensor in terms of 
${\gamma_{ij}}^{kl}$ (denoted by {\tt Gam}):

\tsBrule
\begin{verbatim}DefineTensor[Gam,{{2,1,3,4},1,{1,2,4,3},1}]
DefUnique[Gam[li_,lj_,lk_,ll_],0,PairQ[li,lj]]

RuleUnique[GamRule1,Metricg[li_,lj_] 
      TraceFreeRicciR[lk_,ll_] TraceFreeRicciR[lm_,ln_],
     (- 4 Gam[li,lj,um,uo] 
      + 4 Symmetrize[Metricg[li,um] Metricg[lj,uo],{um,uo}]) 
      TraceFreeRicciR[lm,ln] TraceFreeRicciR[lo,un],
      PairQ[lm,lk]&&PairQ[ll,ln]]
RuleUnique[GamRule2,Metricg[li_,lj_] CD[ScalarR,lk_,ll_],
      (-4 Gam[li,lj,ul,uf]+4 Symmetrize[Metricg[li,ul] 
   Metricg[lj,uf],{ul,uf}]) CD[ScalarR,ll,lf],PairQ[lk,ll]]
Bach3[li_,lj_]=Tsimplify[Dum[Expand[Bach2[li,lj]
                         /.GamRule1/.GamRule2]]]
TraceBach3= Expand[Metricg[ui,uj] Bach3[li,lj]]\end{verbatim}
\tsErule

\section{Gravitational Field Equation of the P\&R Model}

The gravitational field equation (\ref{feq2}) can now be made explicit 
by substituting (\ref{lgeo}), (\ref{fix}), and (\ref{bach}) into it:
\beq \frac{(1+\beta)v^2}{12}\,\frac{1}{\sqrt{-g}}\,\frac{\delta\left(
\sqrt{-g}\,R\right)}
{\delta g_{ij}}+\frac{\lambda v^4}{4}\,\frac{1}{\sqrt{-g}}\,
\frac{\delta\sqrt{-g}}{\delta g_{ij}}+\rho B^{ij}=\frac{1}{2}\,T^{ij}\,.
\label{feq3}
\eeq
We compute (by MathTensor) the variations ($G^{ij}$ = Einstein tensor)
\beq
  \frac{1}{\sqrt{-g}}\,\frac{\delta\left(
  \sqrt{-g}\,R\right)}
  {\delta g_{ij}}=-\, G^{ij}:=- R^{ij} + \frac{1}{2}\,R\,g^{ij}\label{var1}\,
\eeq
and
\beq
\frac{1}{\sqrt{-g}}\,\frac{\delta\sqrt{-g}}{\delta g_{ij}}=
\frac{1}{2}\,g^{ij}\,.\label{var}\eeq
By inserting these relations into (\ref{feq3}), we find the (corrrected) 
P\&R field equation (see \cite[(Eq.\ts 7.2)]{PW2} and \cite[(Eq.\ts 5.2)]{PW})%
\index{P\&R field equation}
\beq
   -\frac{(1+\beta)v^2}{12}\,G^{ij}+\frac{\lambda v^4}{8}\, g^{ij}+
   \rho\, B^{ij}=\frac{1}{2}\,T^{ij}\,.
   \label{feq4}
\eeq 
Provided $\beta\neq -1$ and $v^2\neq 0$, we can put (\ref{feq4})
in a more conventional form,
\beq
G_{ij}-\frac{3\lambda v^2}{2(1+\beta)}\,g_{ij}-\frac{12\rho}{(1+\beta)v^2}
\,B_{ij}=-\frac{6}{(1+\beta)v^2}\,T_{ij}\,,\label{feq5}
\eeq
or, after some (computer) algebra 
($T:=T^k{}_k\,, \;\, \tft_{ij} := T_{ij} - T\, g_{ij}/4$):
\beq 
   R_{ij}+\frac{3\lambda v^2}{2(1+\beta)}\,g_{ij}-
   \frac{12\rho}{(1+\beta)v^2}\,
   B_{ij}=-\frac{6}{(1+\beta)v^2}\,\left( T_{ij}-\frac{1}{2}\,T\,g_{ij}
   \right)\,.\label{feq6}
\eeq
We can decompose this equation into its two irreducible pieces, the tracefree
and the trace piece:
\beq
\left\{
\begin{array}{r c l}
\ds \tfr_{ij}-\frac{12\rho}{(1+\beta)v^2}\,B_{ij} 
& = &
\ds  -\frac{6}{(1+\beta)v^2}\;\tft_{ij}\,,\\[4mm] 
\ds  R+\frac{6\lambda v^2}{(1+\beta)}
& = &\ds   \frac{6}{(1+\beta)v^2}\;T\,.
\end{array}
\right.
\label{feq78}
\eeq 
These two pieces (\ref{feq78}) of the (corrected) 
gravitational field equation (\ref{feq4}) or (\ref{feq6}) of P\&R, together
with the explicit form of the Bach tensor (\ref{bach2}), represent the
{\em general result} of our considerations. 
Note that in the trace piece of (\ref{feq78}) there occur only second
derivatives of the metric, in contrast to the fourth order derivatives
featuring in the Bach tensor of the tracefree piece.
Incidentally, models leading to somewhat similar gravitational field equations
have been discussed since 
the early 1920s by many people (see \cite{fiedler,schim} and 
\cite{perlick}, and the literature cited therein).

In vacuo we have 
\beq
\left\{
\begin{array}{r c l}
\ds \tfr_{ij}&=&
\ds \frac{12\rho}{(1+\beta)v^2}\,B_{ij}\,,\\[4mm] 
\ds R&=&\ds -\frac{6\lambda v^2}{1+\beta} =: 4 \, \Lambda_{\rm cosm}\,.
\end{array}
\right.
\label{feq910}
\eeq 
A glance at (\ref{bach2}) shows that we
can find a {\it special} solution \cite{PW} of the {\it vacuum} equation 
(\ref{feq910}) by using the%
\index{Einstein equation!with cosmological constant}
Einstein vacuum equation with a suitable cosmological constant as an ansatz: 
\beq
   R_{ij}=\Lambda_{\rm cosm}\,g_{ij}\label{eincos}
\eeq or,
alternatively, 
\beq
\left\{
\begin{array}{r c l}
\tfr_{ij}&=&0\,,\\[1mm]
R&=&4\,\Lambda_{\rm cosm}\,.
\end{array}
\right.
\label{eincos'}
\eeq
Then the Bach tensor (\ref{bach2}) vanishes, $B_{ij}=0$, and (\ref{feq910}) 
is fulfilled.

We have collected the MathTensor code that verifies the results of 
this section:

\tsBrule
\begin{verbatim}DefineTensor[B,{{1,2},1}]
DefineTensor[T,{{1,2},1}]
DefineTensor[TrFrT,{{1,2},1}]
DefUnique[B[li_,lj_],0,PairQ[li,lj]]
DefUnique[T[li_,lj_],T,PairQ[li,lj]]
lagHE=Sqrt[-Detg] ScalarR
varHE1=Variation[lagHE,Metricg]
varHE2=PIntegrate[varHE1,Metricg]
HE[ui_,uj_]=Expand[1/Sqrt[-Detg] VariationalDerivative[
            varHE2,Metricg,li,lj]]
lagconst=Sqrt[-Detg]
varconst1=Variation[lagconst,Metricg]
const[ui_,uj_]=Expand[1/Sqrt[-Detg] VariationalDerivative[
               varconst1,Metricg,li,lj]]
MetricgFlag=True

FieldEq1[li_,lj_]=(1/12 (1+beta) v^2 HE[li,lj]) +
                  (1/4 lambda v^4 const[li,lj]) +
                  (rho B[li,lj]) - (1/2 T[li,lj])
FieldEq2[li_,lj_]=Simplify[Collect[Expand[-12/((1+beta) v^2) 
               *FieldEq1[li,lj]],{RicciR[li,lj],ScalarR}]]
TraceFieldEq2=Simplify[Expand[Metricg[ui,uj] 
                              FieldEq2[li,lj]]]
RuleUnique[TRule,T[li_,lj_],
           TrFrT[li,lj]+1/4 T Metricg[li,lj]]
FieldEq3[li_,lj_]=Tsimplify[Expand[FieldEq2[li,lj]-
                  1/2 TraceFieldEq2 Metricg[li,lj]]]
TraceFieldEq3=Expand[Metricg[ui,uj] FieldEq3[li,lj]]
TraceFreeFE[li_,lj_]=Expand[FieldEq3[li,lj] -
                     1/4 Metricg[li,lj] TraceFieldEq3
                     /.RicciToTraceFreeRicciRule/.TRule]\end{verbatim}
\tsErule

\section{Discussion} 
Models very similar in their gravitational sectors to that by P\&R
have been developed, amongst others, by Gregorash \& Papini \cite{Gr3,Gr3a}
and in \cite{KI,PRs}, the latter one, though, in a metric-affine
spacetime with additional conformal invariance \cite[Sect.\ts 6]{PRs}. 
Since conformal invariance is accommodated much more naturally in a 
spacetime with a Weyl piece, we believe that these post-Riemannian
model should be reconsidered in the light of the more recent
developments.
\section*{Acknowledgements}
We are grateful to R.~\Rac\ and M.~\Paw\ for interesting
discussions about their model. 
Furthermore we thank R.\ Schimming for hints to the literature
and F. Gronwald, E.W. Mielke, and Yu. Obukhov for useful remarks. 
One of us (RAP) is supported by the Graduiertenkolleg 
{\em  Scientific Computing}, K\"oln--St.Augustin. 


\end{document}